**ResFrac Blog Post – Uploaded To arXiv**

Commentary on Four New DFIT Papers: (a) Direct In-Situ Measurements of Fracture Opening/Closing from the EGS Collab Project; (b) Comparison of Stress Measurement Techniques from the Bedretto Project; (c) a Statistical Summary of 62 DFITs Interpretations Across Nine Shale Plays; and (d) A Different Perspective: An Article Advocating the Use of the Tangent Method

Mark McClure

mark@resfrac.com

February 7, 2022

This post provides commentary on recent four papers on diagnostic fracture injection testing (DFIT). The first paper uses in-situ deformation measurements to directly observe fractures opening and closing during fracture injection-falloff tests (Guglielmi et al., 2022). The second compares various stress measurement techniques in a series of fracture/injection tests from the Bedretto project (Bröker and Ma, 2022). The third statistically reviews results from applying the interpretation procedure from McClure et al. (2019) to 62 DFITs across nine different shale plays (McClure et al., 2022). The fourth is an op-ed written in JPT (Journal of Petroleum Technology) by an advocate of the tangent method for estimating DFIT closure stress (Buijs, 2021; 2022).

This article presupposes that the reader already has familiarity with these topics. If you would like more background, please refer to McClure et al. (2019).

**Direct In-situ Measurements of Fracture Closure/Reopening from Guglielmi et al. (2022)**

A new paper from Guglielmi et al. (2022) uses direct, in-situ strain measurements to evaluate the accuracy of different methods for stress estimation. Theoretical predictions and computational modeling have suggested that a commonly-used method of estimating stress – the 'tangent method' from Barree et al. (2009) – is often inaccurate (McClure et al., 2014; 2016; 2019; McClure, 2017). McClure et al. (2014; 2016; 2019) proposed an alternative procedure, the 'compliance method.' To test this prediction, there is a need for direct, in-situ measurements that can be compared against different proposed stress estimation techniques to assess accuracy.

Guglielmi et al. (2022) describe fracture injection-falloff tests performed in two different wellbores at the Sanford Underground Research Facility as part of the EGS Collab project (https://eesa.lbl.gov/projects/the-egs-collab-project/). Consistent with the theoretical predictions, the results show that the tangent method interpretations are lower than the actual closure/reopening pressure measured by the strain gauges.

In three of the four transients, the tangent method interpretation is that the fracture never closed during the entire shut-in period, even though the strain measurements demonstrate closure during the shut-in. In the fourth transient, the tangent and compliance estimates are within 100 psi; in this case, the tangent method estimate is consistent with the estimate from the strain gauges.

In contrast, the compliance method interpretations are consistent with the strain measurements in all four cases (with some ambiguity around fracture geometry in the third test, as discussed in the paper).

These findings follow on the heels of another recent study – Dutler et al. (2020) – in which in-situ strain gauges were used to measure fracture closure/reopening at the Grimsel Test Site (https://www.grimsel.com/). As with Guglielmi et al. (2022), repeated measurements were taken in different wellbore intervals, and the tangent method interpretations were usually too low, and often far too low.

These findings are also consistent with results from the M-site tiltmeter data (Branagan et al., 1996), where the compliance method interpretation correctly predicts the reopening pressure measured from the tilt and the tangent method interpretation is too low (Figures 14-15 from McClure et al., 2017; Figure 9A-1 from Gulrajani and Nolte, 2000).

Craig et al. (2017) provide an alternative interpretation of the M-site tiltmeter measurements from Branagan et al. (1996) and arrive at a lower stress estimate that is closer to the tangent method interpretation. It is illuminating to compare the interpretations and explain why they differ.

The figure below, taken from Guglielmi et al. (2022), shows measured in-situ displacement plotted versus measured pressure. This figure is closely based on Figure 9A-1 from Gulrajani and Nolte (2000).

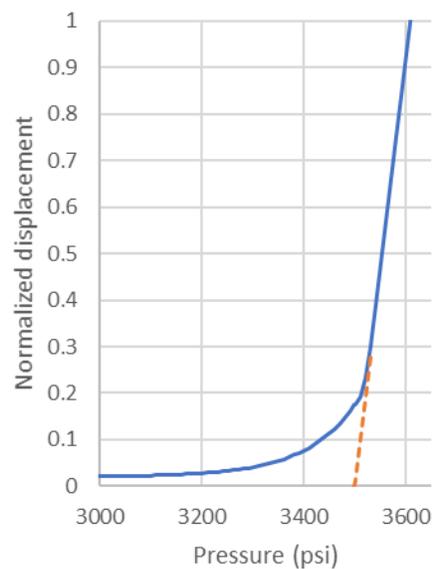

When the fracture walls are out of contact, there is an approximately linear relationship between displacement and pressure, as given by conventional linear elastic fracture mechanics (Sneddon, 1946; Equation 9-21 from Gulrajani and Nolte, 2000; Section 8.13 from Jaeger et al., 2007):

$$\overline{W} = \frac{P - Shmin}{S_f} = (P - Shmin)C_f$$

Where $\overline{W}$ is average aperture, $P$ is pressure, Shmin is the minimum principal stress, $S_f$ is the crack stiffness, and $C_f$ is the fracture compliance.

When the fracture walls come in/out of contact, the slope deviates from the straight line, as the curve bends towards an asymptotic limit near zero the pressure decreases. The bending of the curve occurs as the fracture stiffness increases due to the contacting of the fracture walls (Barton et al., 1985). The deviation from linearity occurs at the point when the fracture walls come back into contact. The extrapolation of the straight line is equal to the magnitude of the minimum principal stress (Gulrajani and Nolte, 2000). This can be seen from examination of the fracture width equation provided above. In the equation, $\overline{W}$ is equal to zero when pressure is equal to Shmin. Thus, the extrapolation of the linear portion of the curve to the x-axis corresponds with the magnitude of Shmin. If the fracture walls were perfectly smooth, then the curve would remain linear all the way until reaching the x-axis. The deviation from the straight line at nonzero $\overline{W}$ occurs because of fracture roughness, which is why the minimum principal stress is slightly lower than the 'contact pressure.'

In contrast, Craig et al. (2017) argue that the stress estimate should be at a lower pressure – a point where the displacement has increased by an appreciable (somewhat arbitrarily defined) amount above zero. However, as shown above, 'the point where displacement is greater than zero' may be much lower than Shmin. The difference between this point and Shmin depends on the fracture roughness and the stiffness of the fracture's asperities. These properties are formation specific and difficult to predict in advance.

Further, there is no theoretical reason to expect the tangent method to identify the point where displacement is appreciably greater than zero. For example, Figures 10 and 13 from Guglielmi et al. (2022) show examples where tangent method interpretations are even lower than the point where strain measurements have reached an approximate minimum. From a theoretical perspective, all we can predict is that the tangent method tends to yield a value lower than the true magnitude of Shmin; sometimes this may be coincidentally in the vicinity of the point where displacement becomes appreciably greater than zero, and sometimes it may not.

To conclude – all three published examples with direct strain measurements show that the tangent method tends to underestimate Shmin (McClure et al., 2016; Dutler et al., 2020; Guglielmi et al., 2022), which is consistent with theoretical predictions (McClure et al., 2016; 2019).

**Comparison of stress interpretation methods at the Bedretto project**

Bröker and Ma (2022) performed a series of fracture injection tests and compare a variety of stress estimation techniques: the compliance method, the tangent method, and others. They found that the tangent method yielded "by far" the lowest estimates of the methods.

There is a physical limit on how low the minimum principal stress can be. If it is too low, the rock ceases to be in 'frictional equilibrium.' Under such conditions, the rock breaks and/or preexisting fractures slide. This failure causes the stress to return back to an equilibrium state. This principal has been established as setting a limit on the stress state in the Earth, based on decades of stress mapping at thousands of sites across the world (Townend and Zoback, 2000; Zoback, 2007).

Bröker and Ma (2022) found that in many cases, the tangent method estimates were so low that they violated frictional equilibrium, even using the most conservative possible assumptions. The authors concluded "The tangent method underestimates the stress magnitude and yields estimates that are below the frictional limit of the rock mass."

The Bröker and Ma (2022) transients had mostly monotonically decreasing dP/dG, consistent with the 'rapid closure' interpretation (McClure et al., 2019). The monotonic dP/dG topic is discussed more by McClure et al. (2019; 2022).

**Statistical review of 62 DFIT Interpretations**

This January, I presented SPE-205297 (McClure et al., 2022) at the International Hydraulic Fracturing Technology Conference in Muscat. In the paper, we statistically review 62 DFIT analyses that we performed for ten different companies across nine different shale plays.

As we prepared this paper in the second half of 2020, we approached every operator for whom we had performed DFIT interpretations and asked for permission to include their anonymized and aggregated DFITs in the study. Nearly all gave permission, and several also gave permission to provide actual DFIT plots in the paper. The study's dataset includes every DFIT from the companies that granted us permission. This study design was designed to avoid the potential for selection bias.

A huge thanks to the companies who gave permission to include their data in the study. A full list is provided in the acknowledgments of the paper.

I am excited about this paper because it 'closes the loop' on the DFIT journey that I and colleagues started eight years ago. We started with theoretical predictions – based in mathematical solutions to fracture mechanics equations – that suggested the tangent method should be inaccurate (McClure et al., 2014; 2016). As discussed above, those predictions have subsequently been confirmed by direct, in-situ strain measurements. Then, in 2018, I organized a collaborative industry study, in which we assembled experts from a group of operators, and over the course of the year, we built-out a step-by-step interpretation procedure. The results were presented in the paper URTeC-2019-123 (McClure et al., 2019).

The McClure et al. (2019) procedure was motivated by the realization that you cannot simply swap out the tangent method and then continue using the rest of the Barree et al. (2009) procedure. The Barree et al. (2009) procedure leads to several inaccuracies, not only for stress, but also for permeability. Some of the procedure's equations use unrealistic assumptions that effectively 'cancel each other out,' and so in some cases, swapping out the tangent method for a more accurate closure pick may *degrade* the accuracy of the calculation (such as the equation for frac efficiency, as discussed in Section A.12 of McClure et al., 2019).

So, we needed to build a new interpretation procedure from the bottom up. Not only were we replacing the use of the 'tangent' method, we needed procedures for estimating permeability, frac efficiency, and other parameters.

Even though I am calling our procedure 'new,' it is deeply rooted in classical stress estimation guidance from Nolte (1979) and Castillo (1987), and permeability/leakoff coefficient estimation methods from Nolte (1979), Mayerhofer et al. (1995), Valko and Economides (1999), Gulrajani and Nolte (2000), Soliman et al. (2005), and Craig and Blasingame (2007). We synthesized these approaches, added our own incremental improvements, included some new ideas, and provided guidance on how to navigate nuances particular to applications in horizontal wells in shale.

In our new paper (McClure et al., 2022), we summarize the results from deploying this procedure across a large, diverse population of DFITs. To quote the abstract:

"We find that: (1) a 'compliance method' closure signature is apparent in the significant majority of DFITs; (2) in horizontal wells, early time pressure drop due to near-wellbore/midfield tortuosity is substantial and varies greatly, from 500 to 6000+ psi; (3) in vertical wells, early-time pressure drop is far weaker; this supports the interpretation that early-time pressure drop in horizontal wells is caused by near-wellbore/midfield tortuosity from transverse fracture propagation; (4) the (not recommended) tangent method of estimating closure yields Shmin estimates that are 100-1000+ psi lower than the estimate from the (recommended) compliance method; the implied net pressure values are 2.5x higher on average and up to 5-6x higher; (5) as predicted by theory, the difference between the tangent and compliance stress and net pressure estimates increases in formations with greater difference between Shmin and pore pressure; (6) the h-function and G-function methods allow permeability to be estimated from truncated data that never reaches late-time impulse flow; comparison shows that they give results that are close to the permeability estimates from impulse linear flow; (7) false radial flow signatures occur in the significant majority of gas shale DFITs, and are rare in oil shale DFITs; (8) if false radial signatures are used to estimate permeability, they tend to overestimate permeability, often by 100x or more; (9) the holistic-method permeability correlation overestimates permeability by 10-1000x; (10) in tests that do not reach late-time impulse transients, it is reasonable to make an approximate pore pressure estimate by extrapolating the pressure from the peak in t*dP/dt using a scaling of $t^{-1/2}$ in oil shales and $t^{3/4}$ in gas shales. The findings have direct practical implications for operators. Accurate permeability estimates are needed for calculating effective fracture length and for optimizing well spacing and frac design. Accurate stress estimation is fundamental to hydraulic fracture design and other geomechanics applications."

Overall, our new paper documents the concordance of theory with observation. The interpretation procedure from McClure et al. (2019) was developed by a group of experts and firmly grounded in analytical derivations and numerical simulations based in mainstream fracture mechanics and reservoir engineering (Valko and Economides, 1995; Gulrajani and Nolte, 2000; Smith and Montgomery, 2015). When applied to real data, we have found that these techniques work, and that everything fits together and behaves as expected.

**A different perspective**

In the January issue of JPT, there is an op-ed by Buijs (2022), which summarizes results from his recent paper, Buijs (2021). Using a series of case studies, the papers argue in favor of using the Barree et al. (2009) DFIT interpretation procedures, in favor of using the tangent method, and specifically *against* using the compliance method.

It's great to see efforts at comparing between methods and seeking independent evidence to validate stress interpretations. Unfortunately, the comparison from Buijs (2021; 2022) was affected by a series of misconceptions about the compliance method. It did not accurately implement the procedure or apply it in an appropriate context. There were also several other major technical problems, as discussed below.

**1.** Buijs (2021; 2022) compare permeability estimates from the holistic permeability equation (Equation A-10 from Barree et al., 2009) and what they call 'compliance method based permeability estimates.' However, they do not actually apply any of the DFIT permeability estimation techniques put forth by developers of the compliance method (McClure et al., 2019; Wang and Sharma, 2019). Instead, the paper applies the Barree et al. (2009) holistic permeability estimation equation with the compliance closure pick, and then calls this a 'compliance based permeability estimate.' This is not an approach that would be recommended by anyone involved in the development of the compliance method. A genuine comparison of the methods would have used the procedures from McClure et al. (2019) or Wang and Sharma (2019).

In a statistical comparison across ten shale plays, McClure et al. (2022) found that Equation A-10 from Barree et al. (2009) yields values 3-1000 times greater than compliance method estimates. Yet, because Buijs (2021) does not actually apply compliance method permeability estimates, he inaccurately states that compliance-based techniques yield *higher* permeability estimates, instead of lower estimates.

Fowler et al. (2019) steps through how permeability overestimates, such as from the equation advocated by Buijs (2021, 2022), lead to suboptimal fracture design and well spacing. The implications of overestimated permeability are apparent in the unrealistically short estimates (10s of feet) for effective fracture length in shale from publications such Barree et al. (2015).

**2.** According to the estimates from Buijs (2021), most of his case studies are from formations with permeability of 0.1-1 md. If these estimates are accurate (which is uncertain), these aren't appropriate datasets for comparison between the methods.

McClure et al. (2019) says "[this] study focused primarily on DFIT's performed in formations with permeability on the order of nanodarcies to 10s of microdarcies. With higher permeability, transients behave qualitatively differently, and many of the techniques in this paper cannot be applied (discussed in Section 4.1)." This is also discussed on pages 1330-1331 from McClure et al. (2016).

**3.** Buijs (2021; 2022) does not properly apply compliance method guidance for estimating stress.

Several times, Buijs (2021) cites 'net pressure' values less than 75 psi. This is impossible because McClure et al. (2019) recommend estimating stress as 75 psi less than the pressure at the time identified as 'fracture contact.'

Also, Buijs (2021) is not appropriately handling tests with monotonic dP/dG. Figure 4.A from Buijs (2021) shows a test with monotonically decreasing dP/dG. As discussed in on pages 1330-1331 from McClure et al. (2016) and in Section 4.1 from McClure et al. (2019), the typical compliance pick is not possible in such tests. In a horizontal well, McClure et al. (2019) recommends against making any interpretation of stress (also discussed in Section 2.2.1 from McClure et al., 2022). In a vertical well, McClure et al. (2019)

says the most-likely interpretation is that the fracture closed soon after shut-in. The paper does not provide specific instructions on exactly how to estimate stress in that context, other than saying that the stress is probably close to the ISIP. McClure et al. (2022) says "in this case, we may cautiously interpret that the fracture closed rapidly and Shmin is close to the ISIP."

Buijs (2021; 2022) does not mention these nuances and instead picks the 'compliance' closure in Figure 4.A from a small wiggle in the G*dP/dG plot. This pick does not actually use the standard compliance method. When it is compared with the guidance from McClure et al. (2019), the pick actually is in the ballpark of where we would have estimated stress – 'close to the ISIP' (note, this is a vertical well). However, because of the monotonic dP/dG, we would express this as a stress estimate with relatively high uncertainty - a few hundred psi. This uncertainty is material because, as discussed below, the tangent estimate is only slightly lower, and Buijs (2021; 2022) uses the picks to make arguments based on stress differences of similar magnitude as these uncertainty ranges.

**4.** Based on the case study, Buijs (2022) makes two arguments based on calculations from 1D stress logs. The first argument, related to 'matching multiple stress measurements,' is not clearly explained, but my best understanding is they felt that the stress log calculations calibrated to the tangent method were more accurate than those calibrated from the compliance method. The second argument is related to a stress log prediction that fracturing will transition to be horizontal at shallower depth.

In this particular Buijs (2022) dataset, the stress estimates from the tangent and compliance methods are unusually close together, within 100-200 psi (Buijs, 2022). This makes it a bad example for comparison, because the methods yield estimates that are so close that they will be difficult to differentiate.

There are a variety of methods for calculating stress logs, and they give different results (Blanton and Olson, 1999; Higgins et al., 2008; Ma and Zoback, 2020). These methods have significant uncertainty because they are derived from highly simplifying/heuristic assumptions (Blanton and Olson, 1999) and because the construction of the logs relies on approximate correlations from well logs. Ma and Zoback (2020) question the underlying premise used by conventional stress log equations and argue for an entirely different approach.

Because of all these approximations and uncertainties, stress logs provide nothing more than practical engineering estimates – they are imprecise and sometimes unreliable. Even if the compliance interpretations were performed correctly (and as discussed above, they were not), such stress log interpretations would not provide 'hard proof' of anything – especially when weighed against direct physical measurements from studies like Dutler et al. (2020) and Guglielmi et al. (2022). This is especially true because the compliance and tangent methods are so close-together in this particular dataset, and because the specifics of the analysis are not clearly provided in the manuscript text.

**5.** On pages 57-58, Buijs (2021) claims that the lower net pressure estimates from the compliance method result in more height growth. This is inconsistent with fracture mechanics. Fractures with greater net pressure build up greater stress intensity factor at the tip and have more energy driving them to propagate. This allows them to propagate further into higher stress zones, and they have an easier time propagating through stress barriers (Figure 6-9 from Mack and Warpinski, 2000).

The proprietary fracture simulator used by Buijs (2021; 2022) uses several adjustments to avoid the unrealistic height growth predictions that would otherwise result from the very high net pressure estimates that often arise from the tangent method (500-1000+ psi, as shown in the review by McClure et al., 2022). These adjustments are: (1) nonstandard methods of calculating fracture compliance and fracture-to-fracture stress shadowing (which is evident when comparing Equations 1 and 2 from Barree, 2019, with Equation 9-21 from Gulrajani and Nolte, 2000, or Section 8.13 from Jaeger et al., 2007), (2) a 'shear decoupling' factor that is used exclusively in this particular simulator (Barree and Winterfeld, 1998), and (3) assignment of different so-called 'process zone stress' values to different layers as a model tuning parameter (Ramurthy et al., 2009).

Buijs (2021) claims that the relatively lower net pressures from the compliance method yield unrealistically low fracture apertures that would cause screenout. I would offer the opposite view – that the 500-1000+ psi net pressures that often arise from the tangent method yield unrealistically short fracture geometries. Worse, when such high values are applied to simulate multifractured horizontal wells in shale, they yield hugely overestimated fracture-to-fracture stress shadowing and injection pressures. If the net pressure of an *individual crack* was 500-1000+ psi, then the aggregated net pressure of a modern fracturing stage with 7-20 perf clusters would be enormous. The fracture simulator employed by Buijs (2021) avoids these overestimated stress shadowing calculations by employing a fudge factor coefficient that allows users to arbitrarily reduce the magnitude of the calculated stress shadow between the fractures, and also by using a highly simplified (and tunable) methods for calculating stress shadowing between fractures (Barree, 2019).

In the past few years, within our company, we have used parameters corresponding to net pressures of 100-300 psi (per individual fracture) in 30+ fracture simulation modeling studies, many of which were calibrated to very detailed 'science pad' datasets. We find these reasonable net pressures can be used within a fracturing simulator – using mainstream fracture mechanics calculations – to yield realistic fracture geometries, which can be closely calibrated to injection data and high quality, direct in-situ measurements from offset fiber and pressure observations (Fowler et al., 2020; Shahri et al., 2021; Ratcliff et al., 2022). Thus, from my perspective, mainstream fracture mechanics is an extremely successful and explanatory theory.

The unusual modeling assumptions advocated by Barree (2019), such as a stress shadow 'tuning coefficient,' are only necessary because of the need to counteract inaccurate net pressure estimates that arise from the tangent method interpretations, and perhaps also to counteract the effects of other simplifying assumptions used by that particular fracturing simulator. Indeed, the tangent method itself is a sharp deviation from conventional techniques for estimating stress that were used historically in petroleum engineering, and that are used currently in mining engineering, rock mechanics, and geophysics (Castillo, 1987; Zoback, 2007; Schmitt and Haimson, 2017).

**6.** Buijs (2022) refers to a paper by Lai et al. (2016) with hydraulic fracturing laboratory experiments in gelatin. He says that the Lai experiments demonstrate that hydraulic fractures in the subsurface do not close 'tip to center.' This claim relies on the questionable assumption that lab-scale fractures in gelatin have comparable surface roughness to fractures formed from hydraulic fracturing in rock (Branagan et al., 1996; van Dam and de Pater, 2001; Vogler et al., 2018). But more importantly, the compliance method does not rely on the idea that fractures close 'tip to center.' The premise of the compliance method is that there is an increase in stiffness caused by the contacting of the fracture walls. This must occur whether or not the fracture closes 'tip to center' (Appendix A from McClure et al., 2016). As McClure et al. (2016) says: "the much-larger deflection [in the derivative] occurs after complete closure [everywhere along the fracture's face]." McClure et al. (2022) says "the dominant effect on the transient is the changing fracture stiffness after the walls have contacted everywhere across the fracture." Thus, the Buijs (2022) comments regarding the Lai paper suggest a fundamental misunderstanding of the compliance method. Whether or not fractures close tip to center is irrelevant to the theoretical basis of the compliance method.

To summarize, Buijs (2021; 2022) does not properly compare compliance and tangent methods because the work: (a) incorrectly applies the McClure et al. (2019) procedure for stress estimation, (b) presents 'compliance method' permeability estimates that are not actually based on the compliance method techniques, and (c) uses data sets outside of the specified permeability range for the compliance technique.

In addition, Buijs (2021; 2022): (a) make claims inconsistent with fracture mechanics, (b) minimize the uncertainty inherent to stress log correlations, (c) misstate the theoretical basis for the compliance method.

It is impossible to independently evaluate many of the interpretations from Buijs (2021; 2022) because the data is proprietary and incompletely provided in the paper's text. For example, in Case 7, Buijs (2021) estimates that an effective infinite conductivity half-length is only 0.004 m. Is this very low number plausible? This seems questionable, but the reader does not have enough information to reproduce the analysis and fully assess.

Comparison between methods is a great idea. But comparisons should be performed with a collaborative approach, or at least, with a conscientious effort to listen to all perspectives. Otherwise, we may encounter the sorts of misunderstandings and inaccuracies exhibited by Buijs (2021; 2022).

It definitely would be interesting to evaluate and extend the compliance method concepts to higher permeability, and to evaluate the accuracy of the Barree et al. (2009) procedures under this range of conditions. Maybe there is a future industry study for us out there…